\documentclass[twocolumn]{svjour3}          % twocolumn

\smartqed  % flush right qed marks, e.g. at end of proof
\usepackage{graphicx}

\usepackage{amssymb}
\usepackage{amsmath}
\usepackage{amsbsy}
\usepackage{multirow}

\journalname{Computational Mechanics}

\begin{document}

\title{GPU accelerated fast multipole boundary element method for simulation of 3D bubble dynamics in potential flow}


\titlerunning{GPU accelerated FMM BEM for simulation of 3D bubble dynamics in potential flow}

\author{N.A. Gumerov         \and
        Yu.A. Pityuk         \and
        O.A. Abramova        \and
        I.S. Akhatov
}


\institute{N.A. Gumerov  \at
              Institute for Advanced Computer Studies, University of Maryland, 21075, USA  \\
              Tel.: +1(301)405-8210\\
              \email{gumerov@umiacs.umd.edu}
           \and
           Yu.A. Pityuk \at
              Center for Micro and Nanoscale Dynamics of Dispersed Systems, Bashkir State University, 450076, Ufa, Russia \\
              Tel.: +7(347)229-96-70\\
              \email{Pityukyulia@gmail.com}
           \and
           O.A. Abramova \at
              Center for Micro and Nanoscale Dynamics of Dispersed Systems, Bashkir State University, 450076, Ufa, Russia \\
              Tel.: +7(347)229-96-70\\
              \email{olgasolnyshkina@gmail.com}
           \and
           I.S. Akhatov \at
              Skolkovo Institute of Science and Engineering (Skoltech), 143026, Moscow, Russia \\
              Tel.: +7 (495) 280 14 81\\
              \email{i.akhatov@skoltech.ru}
}

\date{Received: date / Accepted: date}

\maketitle

\begin{abstract}
A numerical method for simulation of bubble dynamics in three-dimensional
potential flows is presented. The approach is based on the boundary element
method for the Laplace equation accelerated via the fast multipole method
implemented on a heterogeneous CPU/GPU architecture. For mesh stabilization,
a new smoothing technique using a surface filter is presented. This
technique relies on spherical harmonics expansion of surface functions for
bubbles topologically equivalent to a sphere (or Fourier series for toroidal
bubbles). The method is validated by comparisons with solutions available in
the literature and convergence studies for bubbles in acoustic fields. The
accuracy and performance of the algorithm are discussed. It is demonstrated
that the approach enables simulation of dynamics of bubble clusters with
thousands of bubbles and millions of boundary elements on contemporary
personal workstations. The algorithm is scalable and can be extended to
larger systems.

\end{abstract}

\section*{Acknowledgment}

This study is supported by Skoltech Partnership Program, Russian Science Foundation (Grant No.
18-71-00068), and Fantalgo, LLC (Maryland, USA).

\section{Introduction}

Bubbles are common in nature and in many technological processes \cite%
{Brennen1995} including surface cleaning by ultrasound \cite{Xi2012} and
biomedical applications \cite{Ovenden2017}. Very complex physics of
gas-liquid systems may govern the bubble dynamics since at different
conditions different effects can be dominating (e.g., \cite{Nigmatulin1991}%
). So, it is not surprising that most of studies related to single bubble
dynamics or bubbly liquids, where such effects should be taken into account,
treat bubbles as spherical objects (e.g.,
\cite{Plesset1977,Akhatov1997,Khabeev2009,Lauterborn2010,Parlitz1999,Gumerov2012}).
Simulations of bubble dynamics of arbitrary shape are usually
performed using simpler models, such as the model of incompressible inviscid
liquid and a spatially uniform polytropic gas. In this case, boundary
element methods (BEM) are among the most efficient approaches since they
require only boundary discretization, which can be done using a
substantially smaller number of elements compared to the methods based on
volume discretization to achieve the same accuracy.

The BEM for two-dimensional (or axisymmetric) dynamics of a single bubble
near a solid wall and a free surface was developed and used successfully by
many researchers
\cite{Voinov1975,Blake1987,Best1992,Boulton-Stone1993A,Boulton-Stone1993B,Oguz1990,Oguz1993}. In these references also comparisons with experimental data can be
found. A three-dimensional boundary element method was applied to study the
dynamics of bubbles arising from an underwater explosion or induced by a
laser or a spark (e.g.,\cite{Chahine1992,Chahine1994,Zhang1993,Zhang2001}). The BEM was used for determination of the bubble shape \cite{Gumerov2000}, investigation of bubble self-propulsion in potential flows~\cite{Itkulova2014}, energy dissipation during bubble collapse \cite{Lee2007}%
, and bubble dynamics in Stokes flows~\cite{Pozrikidis2003,Itkulova2013}.

Note that large-scale three-dimensional problems are computationally complex
and resource-intensive. Certainly, there is no way to simulate multiphase
flows consisting of billions of bubbles directly, and either continuum
approaches or various schemes coupling micro-,\\ mezo-, and macroscales can be
found in the literature. However, bubble clusters consisting of hundreds or
thousands bubbles may not be well described by continuum theories or
simplified theories neglecting bubble shape effects. Capabilities for
computation of dynamics of such systems can be important for validation of
multiscale approaches and study of various effects in mezoscales. So the
development of methods for acceleration of direct simulations is critical
and such attempts can be found in the literature (e.g., \cite{Bui2006}).

The approach of the present work relies on the BEM accelerated both via a
scalable algorithm, namely, the fast multipole method (FMM), and utilization
of advanced hardware, namely, graphics processors (GPUs) and multicore CPUs.
The primary computational challenge of the classical BEM is related to
solving of a large dense system of $N$ algebraic equations for each time
step, where $N=MN_{d}$ is the total number of collocation points, $M$ is the
number of bubbles, $N_{d}$ is the number of the collocation points on a
single bubble surface. Indeed, in this case, the cost of the direct solution
is $O\left( N^{3}\right) $. This cost can be reduced to $O\left(
N_{iter}C_{MVP}\right) $ using the iterative methods, where $N_{iter}\ll N$
is the number of iterations, and $C_{MVP}$ is the cost of the matrix-vector
product (MVP). If performed directly the latter value can be estimated as $%
O\left( N^{2}\right) $. The application of the FMM reduces the complexity of
MVP operation to $O(N)$, which results in the total cost of the method $%
O\left( N_{iter}N\right) $.

The FMM was introduced by Rokhlin and Greengard \cite{Greengard1987} and
further developed by these and other researchers (particularly, for the
Laplace equation in three dimensions, e.g., \cite{Cheng1999,Gumerov2008}).
Comparison of efficiency of different methods for the Laplace equation in 3D
can be found \cite{GumerovL2005}. A number of authors considered
acceleration of the BEM using the FMM (e.g., \cite{Nishimura2002,Gumerov2006,Liu2009}). The BEM accelerated via the FFTM (an FMM-type scalable algorithm
combining the single level FMM and the FFT) was successfully used for bubble
dynamics simulations by Bui et al \cite{Bui2006}. While both the FMM and the
FFTM have $O\left( N\right) $ or $O\left( N\log N\right) $ complexity, the
difference can appear in simulations of large systems as the FMM may be more
efficient for highly non-uniform distributions, where skipping of empty
boxes can be essential.

The FMM can be efficiently parallelized \cite{Greengard1990}. The first
implementation of the FMM on graphics processors \cite{Gumerov2008} was
developed further \cite{Hu2011,Hu2012}, where the FMM was
implemented on heterogeneous computing architectures consisting of multicore
CPUs and GPUs. This FMM parallelization strategy for heterogeneous
architectures was successfully used in fluid and molecular dynamics \cite{Itkulova2012,Abramova2013,Abramova2014,Hu2013,Maryin2013}
 and in electro- and magnetostatics \cite{Adelman2017}. A similar
approach is applied in the present study for simulation of bubble dynamics
with millions of boundary elements on personal workstations. It should be
mentioned that there exist different FMM parallelization strategies for
heterogeneous architectures and demonstrations of high-performance
applications for simulation of blood flows, turbulence, etc. (e.g., \cite{Lashuk2009,Rahimian2010,Yokota2013}).

It is well known that BEM-based bubble dynamics codes cannot work correctly
without mesh stabilization and smoothing (e.g., \cite{Zhang2001}, also some
literature review can be found here). Indeed, smooth bubble surface is
provided naturally by surface tension and liquid viscosity and
compressibility. In many cases, the spatial and temporal scales related to
these effects are much smaller compared to the characteristic scales of
bubble dynamics (e.g., the frequency of oscillations and the bubble size).
To achieve natural (physical) smoothing substantially small time steps and
fine surface meshes should be employed. In simplified models, where either
all these effects are neglected (e.g., \cite{Bui2006}), or they have taken
into account only partially (e.g., just the surface tension) artificial
stabilization and smoothing of the surface should be used. A similar
situation can be observed, e.g., in the modeling of shock waves, where the
viscosity provides an extremely thin boundary layer, while for simulation of
shock waves in inviscid media various methods are used to stabilize
computations. In the present study, we propose a novel technique for
smoothing based on shape filters. We also provide details necessary for
development and implementation of a stable and efficient bubble dynamics
code.

The goal of this paper is to present the method and show its performance and
scaling with the problem size. For this purpose, we used several benchmark
problems. To validate the code and compare with the data available in the
literature we used small-scale examples (several bubbles). The performance
and scaling are studied on a somewhat artificial configuration of a
``regular'' bubble cluster. The reason for this is that such clusters can be
easily reproduced by other researchers and they can use the data provided in
this paper for comparisons and validations. It should be noticed that the
code can compute the dynamics of bubbles of different topology, including
toroidal bubbles (and the reader can find expressions for the toroidal shape
filter). However, the description for the handling of the topology change
and peculiarities of such modeling (which can be found elsewhere) brings
unnecessary complication to the presentation and does not contribute to the
demonstration of the performance of the method, which is the main goal. So,
for the clarity of presentation, we limited ourselves with examples for
bubbles topologically equivalent to a sphere. The authors also expect future
publications with more physically interesting cases simulated using the
method presented in this paper.

\section{Statement of the problem}

\subsection{Governing equations}

Consider the dynamics of a cluster consisting of $M$ gas bubbles in an
incompressible inviscid liquid of density $\rho $, which motion is described
by equations

\begin{eqnarray} \label{eq2_1}
&& \rho \frac{d\mathbf{v}}{dt}=-\nabla p+\rho \mathbf{g},\quad \nabla \cdot
\mathbf{v}=0,\quad \frac{d}{dt}=\frac{\partial }{\partial t}+\mathbf{v}\cdot
\nabla ,
\end{eqnarray}%
where $\mathbf{v}$ is the liquid velocity, $p$ the pressure, and $\mathbf{g}$
the gravity acceleration. These equations have a solution in the form of
potential flow,
\begin{eqnarray} \label{eq2_2}
&& \mathbf{v}=\nabla \phi ,
\end{eqnarray}%
where $\phi $ is the velocity potential satisfying the Laplace equation at
any moment of time,
\begin{eqnarray} \label{eq2_3}
&& \nabla ^{2}\phi \left( \mathbf{r},t\right) =0.
\end{eqnarray}%
Spatial integration of Eq.~(\ref{eq2_1}) results in the Cauchy-Lagrange
(unsteady Bernoulli) integral,
\begin{eqnarray}  \label{eq2_4}
&& \frac{\partial \phi }{\partial t}+\frac{1}{2}\left| \nabla \phi \right| ^{2}+%
\frac{p}{\rho }=\mathbf{g\cdot r}+F(t),
\end{eqnarray}%
where $F(t)$ is the integration constant, which should be determined from
the boundary conditions at the infinity. For liquid resting far from the
bubble at pressure $p_{\infty }(t)$, we have%
\begin{eqnarray} \label{eq2_5}
&& \left. \phi \right| _{\left| \mathbf{r}\right| \rightarrow \infty }=0,\quad
\left. p\right| _{\left| \mathbf{r}\right| \rightarrow \infty }=p_{\infty
}\left( t\right) +\rho \mathbf{g\cdot r,\quad }\\
&&F\left( t\right) =p_{\infty }\left( t\right) /\rho . \notag
\end{eqnarray}%
Notably, for time-harmonic acoustic fields considered in this study, $%
p_{\infty }(t)$ is specified as
\begin{eqnarray} \label{eq2_6}
&& p_{\infty }(t)=p_{0}+p_{a}(t),\quad p_{a}(t)=-P_{a}\sin \omega t,
\end{eqnarray}%
where $p_{0}$ is the static pressure and $p_{a}(t)$ is the acoustic pressure
characterized by the amplitude $P_{a}$ and the circular frequency $\omega $.

The total gas-liquid interface $S$ is a union of interfaces of all bubbles, $%
S=S_{1}\cup ...\cup S_{M}$, where $S_{m}$ is the surface of the $m$th
bubble, $m=1,...,M.$ The liquid pressure $p(\mathbf{r},t)$ and the gas
pressure, $p_{gm}\left( t\right) $, on the bubble surface $S_{m}$ are
related as
\begin{eqnarray} \label{eq2_7}
&& p\left( \mathbf{r,}t\right) =p_{gm}\left( t\right) -2\sigma H_{m}\left(
\mathbf{r,}t\right) ,\\
&& \mathbf{r}\in S_{m},\quad m=1,...,M, \notag
\end{eqnarray}%
where $\sigma $ is the surface tension and $H_{m}\left( \mathbf{r,}t\right) $
is the mean surface curvature. The gas pressure depends on the bubble volume
according to the polytropic law,
\begin{eqnarray} \label{eq2_8}
&& p_{gm}\left( t\right) =p_{gm0}\left( \frac{V_{m0}}{V_{m}}\right) ^{\kappa
},\\
&&p_{gm0}=p_{0}+\frac{2\sigma }{a_{m0}},\quad m=1,...,M,  \notag
\end{eqnarray}%
where $\kappa$ is the polytropic exponent (for the isothermal
processes $\kappa =1$ and for the adiabatic processes $\kappa =\gamma _{g}$, where $\gamma _{g}$ is the gas specific
heats ratio), subscript ``0'' refers to the
initial value at $t=0$, $V_{m}$ is the $m$th bubble volume, and $a_{m0}$ is
the effective bubble radius at $t=0$ (assuming that the hydrostatic pressure
gradient has a negligible effect on the initial bubble shape).

Evolution of the velocity potential and the gas-liquid interface is
determined by the dynamic and kinematic conditions,
\begin{eqnarray}\label{eq2_9}
&& \frac{d\phi }{dt}=\frac{1}{2}\left| \mathbf{v}(\mathbf{r,}t)\right| ^{2}-%
\frac{p_{gm}\left( t\right) -2\sigma H_{m}\left( \mathbf{r,}t\right) }{\rho }\\
&& +\mathbf{g\cdot r}+F(t), \quad \mathbf{r}\in S_{m},\quad m=1,...,M, \notag
\end{eqnarray}%

\begin{eqnarray} \label{eq2_10}
&&\frac{d\mathbf{r}}{dt}=\mathbf{v}(\mathbf{r,}t),\quad \mathbf{n}\cdot
\mathbf{v}=\frac{\partial \phi }{\partial n}=q,\quad \mathbf{r}\in S,
\end{eqnarray}%
where $\mathbf{n}(\mathbf{r,}t)$ is the normal to the surface $S$. These
relations close the problem. Indeed, Eq.~(\ref{eq2_10}) propagates the
boundary value of the potential to the next time step. The potential also
determines the tangential velocity as the derivative of this quantity along
the surface. The normal component of the velocity can be found from the
solution of the Dirichlet boundary value problem for the Laplace equation.
As soon as the surface velocity is found the position of the interface can
be updated.

\subsection{Boundary integral equations}

The BEM uses a formulation in terms of boundary integral equations (BIE)
whose solution with boundary conditions provides $\phi (\mathbf{r})$ and $q(\mathbf{r})=\partial\phi(\mathbf{r})/\partial n(\mathbf{r})$ on the
boundary and subsequently determines $\phi (\mathbf{r})$ for external and
boundary domain point $\mathbf{r}$. Using Green's identity the boundary
integral equations for $\phi |_{|\mathbf{r}|\rightarrow \infty }=0$ can be
written in the form
\begin{eqnarray} \label{eq2_11}
&& L\left[ q\right] \left( \mathbf{r}\right) -M\left[ \phi \right] \left(
\mathbf{r}\right) =\left\{
\begin{array}{c}
-\phi \left( \mathbf{r}\right) ,\quad \mathbf{r}\notin S,\quad \mathbf{r}\in
\Omega , \\
-\frac{1}{2}\phi \left( \mathbf{r}\right) ,\quad \mathbf{r}\in S,\quad  \\
0,\quad \mathbf{r}\notin \Omega .
\end{array}%
\right.
\end{eqnarray}%
Here $\Omega $ is the domain occupied by liquid, and $L[q]$ and $M[\phi ]$
are the single and double layer potentials, respectively:

\begin{eqnarray} \label{eq2_12}
&& L\left[ q\right] \left( \mathbf{r}\right) =\int_{S}q\left( \mathbf{r}%
^{\prime }\right) G\left( \mathbf{r},\mathbf{r}^{\prime }\right) dS(\mathbf{r%
}^{\prime }),\\
&& M\left[ \phi \right] \left( \mathbf{r}\right)
=\int_{S}\phi \left( \mathbf{r}\right) \frac{\partial G\left( \mathbf{r},%
\mathbf{r}^{\prime }\right) }{\partial n\left( \mathbf{r}^{\prime }\right) }%
dS(\mathbf{r}^{\prime }),  \notag
\end{eqnarray}%
where $G\left( \mathbf{r},\mathbf{r}^{\prime }\right) $ is the free space
Green's function for the Laplace equation, and $\partial G\left( \mathbf{r},%
\mathbf{r}^{\prime }\right) /\partial n\left( \mathbf{r}^{\prime }\right) $
is its normal derivative for the normal directed outside the bubble%
\begin{eqnarray} \label{eq2_13}
&& G\left( \mathbf{r},\mathbf{r}^{\prime }\right) =\frac{1}{4\pi \left| \mathbf{%
r}-\mathbf{r}^{\prime }\right| },\\
&& \frac{\partial G\left( \mathbf{r},%
\mathbf{r}^{\prime }\right) }{\partial n\left( \mathbf{r}^{\prime }\right) }=%
\frac{\mathbf{n}\left( \mathbf{r}^{\prime }\right) \cdot \left( \mathbf{r}-%
\mathbf{r}^{\prime }\right) }{4\pi \left| \mathbf{r}-\mathbf{r}^{\prime
}\right| ^{3}}.  \notag
\end{eqnarray}

\section{Numerical method}

\subsection{Discretization}

Discretization of the boundary results in an approximation of surface
functions via finite vectors of their surface samples and integral operators
via matrices acting on that vectors. In the present study, the surface $S$
is discretized by a triangular mesh with $N$ vertices $\mathbf{r}_{i}$, $%
i=1,...,N$, which used as the collocation points. For a given set of
collocation points, the quadratures for the single and double layer
integrals can be written in the form

\begin{eqnarray} \label{d1}
&& L\left[ q\right] \left( \mathbf{r}_{i}\right) \approx
\sum_{j=1}^{N}L_{lj}q_{j},\quad M\left[ \phi \right] \left( \mathbf{r}%
_{i}\right) \approx \sum_{j=1}^{N}M_{ij}\phi _{j},\\
&& \phi _{j}=\phi
\left( \mathbf{r}_{j}\right) ,\quad q_{j}=q\left( \mathbf{r}_{j}\right) , \notag
\end{eqnarray}%
where $L_{ij}$ and $M_{ij}$ are the elements of the BEM $N\times N$ matrices
$\mathbf{L}$ and $\mathbf{M}$ representing the surface operators. They can
be found by evaluation of the integrals over the triangles sharing a given
collocation point $\mathbf{r}_{j}$. Discretization~(\ref{d1}) of the
boundary integral equation~(\ref{eq2_11}) results in a linear system%
\begin{eqnarray}  \label{eq3_1}
&& \sum_{j=1}^{N}L_{ij}q_{j}=b_{i},\quad b_{i}=-\frac{1}{2}\phi
_{i}+\sum_{j=1}^{N}M_{ij}\phi _{j},\\
&&  i=1,...,N,  \notag
\end{eqnarray}%
which can also be written in the matrix-vector form%
\begin{eqnarray} \label{eq3_2}
&& \mathbf{Lq} =\mathbf{b,\quad b=}-\frac{1}{2}\boldsymbol{\phi }+\mathbf{M}\boldsymbol{\phi }%
,\quad   \\
&& \mathbf{L} =\left\{ L_{ij}\right\} ,\quad \mathbf{M}=\left\{
M_{ij}\right\} , \notag \\
&& \mathbf{q}=\left\{ q_{j}\right\} ,\quad \mathbf{b}%
=\left\{ b_{i}\right\} ,\quad \boldsymbol{\phi =}\left\{ \phi _{i}\right\} .\quad
\notag
\end{eqnarray}

\subsection{Non-singular integrals}

There exist extensive literature for accurate numerical and analytical
evaluation of the integrals of the Green's function and its derivatives over
triangles (e.g., \cite{Chen02,Adelman2016}). However, efficient use of the
FMM for large $N$ requires numerically inexpensive quadratures and
approximations, which brings forward strategies, such as described
\cite{Gumerov2009}. This scheme is used in the present study for computation
of the non-singular elements,%
\begin{eqnarray}
&& L_{ij} =s_{j}G\left( \mathbf{r}_{i},\mathbf{r}_{j}\right) ,\quad
M_{ij}=s_{j}\frac{\partial G}{\partial n_{j}}\left( \mathbf{r}_{i},\mathbf{r}%
_{j}\right) ,\quad i\neq j,  \label{ns1} \\
&& s_{j} =\frac{1}{3}\sum_{S_{k}\ni \mathbf{r}_{j}}A_{k},\quad \mathbf{n}_{j}=%
\frac{\mathbf{m}_{j}}{\left| \mathbf{m}_{j}\right| },\quad \mathbf{m}%
_{j}=\sum_{S_{k}\ni \mathbf{r}_{j}}\mathbf{N}_{k}A_{k},  \notag
\end{eqnarray}%
where $s_{j}$ and $\mathbf{n}_{j}$ are the surface area (weight) and the
unit normal associated with the $j$th vertex, and the summation is taken
over all triangles $S_{k}$ of area $A_{k}$ and normal $\mathbf{N}_{k}$
sharing the vertex. This scheme was compared with higher order quadratures %
\cite{Gumerov2009} and tested on large scale problems for the Helmholtz
equation. It showed good results for ``good'' meshes (a ``good'' mesh
consists of ``good'' triangles of approximately the same size; the goodness
of a triangle is characterized by its deviation from a ``perfect'' triangle,
which is an equilateral triangle).

\subsection{Singular integrals}

Singular integrals can be computed based on the integral identities, which
provide expressions for these integrals via the sums of the regular
integrals over the surface. The identities can be derived from Green's
identities applied to analytical solutions of the test problems. Such
methods were developed and tested for the Laplace and Helmholtz equations
and used by several authors (e.g., \cite{Gumerov2009,Klaseboer2009}). The
method used in the present study is the following.

A test function $\Phi $, which is harmonic and regular inside the interior
of domain $\mathbb{R}^{3}\backslash \Omega $ (inside the bubbles), satisfies
the identity
\begin{eqnarray} \label{s1}
&& \frac{1}{2}\Phi \left( \mathbf{r}\right) =L\left[ Q\right] \left( \mathbf{r}%
\right) -M\left[ \Phi \right] \left( \mathbf{r}\right) ,\quad \mathbf{r}\in
S,\\
&& Q=\frac{\partial \Phi }{\partial n}.  \notag
\end{eqnarray}%
A discrete form of this relation can be written as%
\begin{eqnarray} \label{s2}
&& L_{ii}Q_{i}=\sum_{j\neq i}\left( M_{ij}\Phi _{j}-L_{ij}Q_{j}\right) +\left(
\frac{1}{2}+M_{ii}\right) \Phi _{i},\\
&& i=1,...,N.  \notag
\end{eqnarray}%
\qquad

First, we determine the diagonal elements of matrix $\mathbf{M}$. A
non-trivial regular harmonic function can be taken as $\Phi \left( \mathbf{r}%
\right) \equiv 1$, for which $Q=0$. So requesting that this solution is
exact for the discrete form, i.e., setting $\Phi _{i}=1$ and $Q_{i}=0$ in
Eq. (\ref{s2}), we obtain%
\begin{eqnarray} \label{s3}
&& M_{ii}=-\frac{1}{2}-\sum_{j\neq i}M_{ij},\quad i=1,...,N.  \notag
\end{eqnarray}

To determine the singular elements of matrix $L$ we use three test functions
$\Phi _{1}\left( \mathbf{r}\right) \equiv x,$ $\Phi _{2}\left( \mathbf{r}%
\right) \equiv y,$ $\Phi _{3}\left( \mathbf{r}\right) \equiv z$, which
normal derivatives on the surface are the components of the normal vector $%
\mathbf{n=}\left( n_{x},n_{y},n_{z}\right) $, i.e., $Q_{1}\left( \mathbf{r}%
\right) \equiv n_{x}\left( \mathbf{r}\right) ,$ $Q_{2}\left( \mathbf{r}%
\right) \equiv n_{y}\left( \mathbf{r}\right) ,$ $Q_{3}\left( \mathbf{r}%
\right) \equiv n_{z}\left( \mathbf{r}\right) $. Note then that Eq. (\ref{s2}%
) can be written in the vector form,%
\begin{eqnarray} \label{s4}
&& L_{ii}\mathbf{n}_{i}=\sum_{j\neq i}\left( M_{ij}\mathbf{r}_{j}-L_{ij}\mathbf{%
n}_{j}\right) +\left( \frac{1}{2}+M_{ii}\right) \mathbf{r}_{i},\\
&& i=1,...,N.  \notag
\end{eqnarray}%
Taking the scalar product of this relation with $\mathbf{n}_{i}$ for each
collocation point, we obtain%
\begin{eqnarray} \label{s5}
&& L_{ii}=\mathbf{n}_{i}\left[ \sum_{j\neq i}\left( M_{ij}\mathbf{r}%
_{j}-L_{ij}\mathbf{n}_{j}\right) +\left( \frac{1}{2}+M_{ii}\right) \mathbf{r}%
_{i}\right] ,\\
&& i=1,...,N.  \notag
\end{eqnarray}

The computational cost of the above procedure when using the FMM is equal to
the cost of four FMM function calls (one for the diagonal of $\mathbf{M}$
matrix and three for the diagonal of $\mathbf{L}$ matrix) since a single
call of the FMM can handle input as a sum of monopoles and dipoles. As it is
mentioned below, the number of the FMM calls per time step can be several
times larger. Hence, the method described is consistent with the overall
algorithm complexity. However, it may create 20-30\% overhead for an
FMM-based linear system solver and more efficient methods can be developed
in future.

\subsection{Tangential velocity}

Determination of the full velocity $\mathbf{v}$ for potential flow is needed
for computation of the pressure and time evolution of the surface potential
(see Eqs~(\ref{eq2_9}) and (\ref{eq2_10})). In the present study, we
implemented the following method of surface differentiation consistent with
the low order BEM (constant panel or linear approximations).

The velocity on the surface can be decomposed into its normal and tangential
components,%
\begin{eqnarray} \label{eq3_6}
&& \mathbf{v=n}q+\mathbf{v}_{t},\quad \mathbf{v}_{t}=\left( \mathbf{n}\times
\mathbf{v}\right) \times \mathbf{n}.
\end{eqnarray}%
To obtain $\mathbf{n}\times \mathbf{v}$ we use the Stokes theorem in the
form
\begin{eqnarray}  \label{eq3_7}
&& \int_{S_{k}}\left( \mathbf{n}\times \mathbf{v}\right) dS=\int_{C_{k}}\phi
\left( \mathbf{r}\right) d\mathbf{r,\quad v}=\nabla \phi ,
\end{eqnarray}%
where $S_{k}$ is the $k$th surface triangle and $C_{k}$ is the contour
bounding $S_{k}$. Assume that the triangle has positive orientation for the
path $\mathbf{r}_{k1}\rightarrow \mathbf{r}_{k2}\rightarrow \mathbf{r}%
_{k3}\rightarrow \mathbf{r}_{k1}$ connecting the respective triangle
vertices. At these vertices, the values of $\phi $ are known and can be
denoted as $\phi _{k1},\phi _{k2},$ and $\phi _{k3}$, respectively. The
linear interpolation along the segment $C_{kij}$ connecting $\mathbf{r}_{ki}$
and $\mathbf{r}_{kj}$ can be written in the form%
\begin{eqnarray} \label{eq3_8}
&& \phi \left( \mathbf{r}\right) =\left( 1-\xi \right) \phi _{ki}+\xi \phi
_{kj},\\
&&  \mathbf{r}=\left( 1-\xi \right) \mathbf{r}_{ki}+\xi \mathbf{r}%
_{kj},\quad \xi \in \left[ 0,1\right] ,\quad i,j=1,2,3.  \notag
\end{eqnarray}%
Hence, we have for the line integral along $C_{kij}$
\begin{eqnarray} \label{eq3_9}
&& \mathbf{I}_{kij}=\int_{C_{kij}}\phi (\mathbf{r})d\mathbf{r} \\
&& =\left( \mathbf{r}_{kj}-\mathbf{r}_{ki}\right) \int_{0}^{1}\left[ \left( 1-\xi \right) \phi
_{ki}+\xi \phi _{kj}\right] d\xi  \notag \\
&& =\frac{1}{2}\left( \mathbf{r}_{kj}-\mathbf{r%
}_{ki}\right) \left( \phi _{ki}+\phi _{kj}\right). \notag
\end{eqnarray}%
The surface average value of vector $\mathbf{n}\times \mathbf{v}$ over the
triangle $S_{k}$ can be computed according to Eqs~(\ref{eq3_6}) and (\ref%
{eq3_9}),%
\begin{eqnarray} \label{eq3_10}
&& \left( \mathbf{n}\times \mathbf{v}\right) _{k} =\frac{1}{A_{k}} \int_{S_{k}}\left( \mathbf{n}\times \mathbf{v}\right) dS\\
&& =\frac{1}{A_{k}} \left( \mathbf{I}_{k12}+\mathbf{I}_{k23}+\mathbf{I}_{k31}\right) \notag \\
&& =\frac{1}{2A_{k}}\left[ \left( \mathbf{r}_{k2}-\mathbf{r}_{k3}\right) \phi
_{k1}+\left( \mathbf{r}_{k3}-\mathbf{r}_{k1}\right) \phi _{k2}\right.\notag \\
&& \left.+\left(\mathbf{r}_{k1}-\mathbf{r}_{k2}\right) \phi _{k3}\right] .  \notag
\end{eqnarray}

The value of $\mathbf{n}\times \mathbf{v}$ at the $j$th vertex then can be
computed as an area-based average, similarly to Eq.~(\ref{ns1}). So, the
tangential velocity at the vertex can be found according to Eq.~(\ref{eq3_6}%
),%
\begin{eqnarray} \label{eq3_11}
&& \mathbf{v}_{tj}=\left( \mathbf{n}\times \mathbf{v}\right) _{j}\times \mathbf{%
n}_{j},\\
&& \left( \mathbf{n}\times \mathbf{v}\right) _{j} =\frac{1}{3s_{j}}\sum_{S_{k}\ni \mathbf{r}_{j}}\left( \mathbf{n}\times \mathbf{v}\right)
_{k}A_{k}.    \notag
\end{eqnarray}

\subsection{Surface curvature}

The mean surface curvature $H$ can be computed by the algorithms of contour
integration and fitted paraboloid proposed and discussed in details~\cite{Zinchenko1997}. Both methods were implemented in the present study and
compared. It was found that the contour integration method is more efficient
for coarse meshes, while the fitted \\paraboloid method is more accurate
for higher discretizations. For a good quality mesh ($N_{d}>600$) the
relative errors in the mean curvature computed with the latter method do not
exceed 1\%. We also obtained excellent preliminary results for computation
of the surface curvature using shape filtering technique described below.
However, a more detailed study is needed for this technique, which may be
reported in a separate publication. In the present paper, we used the fitted
paraboloid method (slightly different from the original \cite{Zinchenko1997}%
), which briefly is the following.

For a mesh vertex $\mathbf{r}_{j}$ we use only five neighbor vertices $%
\mathbf{r}_{j}^{(1)},...,\mathbf{r}_{j}^{(5)}.$ The meshes used in the
present computations have vertex valencies at least 5 (the valency is the
number of the edges sharing the same vertex). So, when more than 5 neighbors
are available, we use just 5 (one can use least squares for overdetermined
systems with more than 5 neighbors, but we found that the final result is
not affected substantially by the accepted simplification). We solved then a
linear system of 5 equations to get 5 coefficients of the fitted paraboloid,
$B_{j}^{(1)},...,B_{j}^{(5)}$, for each vertex%
\begin{eqnarray} \label{parab1}
&& \sum_{i=1}^{5}Q_{j}^{(li)}B_{j}^{(i)}=z_{j}^{(l)},\quad l=1,...,5,\quad
j=1,...,N,  \\
&& Q_{j}^{(l1)}=x_{j}^{(l)},\quad Q_{j}^{(l2)}=y_{j}^{(l)},\quad
Q_{j}^{(l3)}=x_{j}^{(l)2}, \notag\\
&& Q_{j}^{(l4)}=x_{j}^{(l)}y_{j}^{(l)},\quad
Q_{j}^{(l5)}=y_{j}^{(l)2},  \notag
\end{eqnarray}%
where $\left( x_{j}^{(l)},y_{j}^{(l)},z_{j}^{(l)}\right) $ are the Cartesian
coordinates of vertex $\mathbf{r}_{j}^{(l)}$ in the reference frame centered
at $\mathbf{r}_{j}$, which $z$ axis has the same direction as the normal $\mathbf{n}_{j}$. An analytical solution of the 5$\times $5 system using
Cramer's rule is implemented in the code. The mean curvature at the $j$th
vertex then computed~as%
\begin{equation} \label{parab2}
H_{j}=-B_{j}^{(3)}-B_{j}^{(5)},\quad j=1,...,N.
\end{equation}

Note that Zinchenko et al~\cite{Zinchenko1997} proposed an iterative process to
update the normal direction based on the coefficients of the fitted
paraboloid, but this was not used in the present implementation. In fact,
one can fit only three coefficients $B_{j}^{(3)},B_{j}^{(4)},$ and $%
B_{j}^{(5)}$, since coefficients $B_{j}^{(1)}$ and $B_{j}^{(2)}$ should be
zero if the normal to the surface coincides with the axis of the paraboloid
(Eq. (\ref{parab2}) neglects terms $O\left( B_{j}^{(1)2}\right) $, $O\left(
B_{j}^{(2)2}\right) $, and $O\left( B_{j}^{(1)}B_{j}^{(2)}\right) $). In
this case, only three neighbors for vertex $j$ are needed. However, there
can appear situations when the 3$\times $3 system is close to degenerate
(e.g., due to symmetry $x_{j}^{(2)}=-x_{j}^{(1)}$, $y_{j}^{(2)}=-y_{j}^{(1)}$%
, $z_{j}^{(2)}=$ $z_{j}^{(1)}$) and some special treatment is required for
such cases.

\subsection{Time marching}

For time integration we tried several explicit schemes, including multistep
methods, such as the Adams-\\Bashforth (AB) schemes of the 1st - 6th orders
and the Adams-Bashforth-Moulton (ABM) predictor-corrector scheme of the 4th
- 6th order. These methods require one call (AB) and two calls (ABM) of the
right-hand side function per time step and data from several previous time
steps. For the initialization, or warming up, we used the Runge-Kutta (RK)
methods of the 4th - 6th orders. After some optimization study, we chose the
6th order AB method warmed up by the 4th order RK method.

The time step used in the explicit schemes~(\ref{eq2_9}) and (\ref{eq2_10})
should be sufficiently small to satisfy a Courant-type stability condition
\begin{eqnarray} \label{eq3_12}
&& \triangle t=C\min (\triangle _{d0})\left( \frac{\rho }{P_{a}}\right)
^{1/2}<\\
&& \min (\triangle _{d}\left( t\right) )\left( \frac{\rho }{P_{a}}%
\right) ^{1/2},  \notag
\end{eqnarray}%
where $\min (\triangle _{d}\left( t\right) )$ is the minimum spatial
discretization length (the length of the edge of the mesh) at the moment $t$%
, $(P_{a}/\rho )^{1/2}$ is the characteristic velocity of bubble
growth/collapse. For integration with a constant time step, one can use $%
C\min (\triangle _{d0})$ as the lower bound of $\triangle _{d}\left(
t\right) $, where $\triangle _{d0}$ is the discretization length at $t=0$,
and $C$ is some constant. This constant can be found empirically based on
the particular problem (normally $C\ll 1$). It is also noticeable that the
current method of solving is iterative at each time step, where the initial
guess is taken from the previous time step. Even though one can select $%
C\sim 1$ and have a stable integration, the number of iterations per time
step may substantially increase if $C$ is not small. So, the selection of $C$
is a subject for optimization, where the overall integration time can serve
as an objective function. For the examples reported in the present study, we
found that reasonable values of $C$ are of the order of $0.1.$

For mesh stabilization, the surface points can be forced to move with some
velocity $\mathbf{u}$, which is different from the liquid velocity,
\begin{eqnarray} \label{eq3_13}
&& \mathbf{u=v}+\alpha \mathbf{v}_{t}.
\end{eqnarray}%
Here $\alpha $ is some correction factor. Particularly, at $\alpha =-1$ we
have $\mathbf{u=n}q$ (see Eq. (\ref{eq3_6})), and this value is used in the
present simulations.

So, a surface point moves according to
\begin{eqnarray} \label{eq3_14}
&& \frac{d\mathbf{r}}{dt}=\mathbf{u,\quad r}\in S,
\end{eqnarray}%
while the potential at this point evolves as
\begin{eqnarray} \label{eq3_15}
&& \frac{d\phi }{dt}=\frac{\partial \phi }{\partial t}+\mathbf{u}\cdot \nabla
\phi =\left( \alpha \mathbf{v}_{t}+\frac{1}{2}\mathbf{v}\right) \cdot
\mathbf{v} \notag \\
&&-\frac{p_{gm}\left( t\right)-2\sigma H_{m}\left( \mathbf{r,}t\right) }{\rho }+\mathbf{g\cdot r+}F(t),   \\
&& \quad \mathbf{r}\in S_{m},\quad m=1,...,M.  \notag
\end{eqnarray}%
Another essential technique we use to stabilize the mesh is the shape filter
described below.

\subsection{Surface/shape filter}

Traditional boundary element methods suffer from some geometric errors
related to flat panel representation of the surface (errors in computation
of surface integrals, normals, areas, and tangential components), which
result in destabilization of the mesh in dynamic problems (appearance of the
noise and the mesh jamming). The use of approximate methods for solving of
linear systems (such as iterative methods with approximate matrix-vector
multiplication) also destabilizes the mesh. A noisy surface can be smoothed
out using some bandlimited parametric representation of the surface. The
shape filter is a linear operator, which takes as input coordinates of
surface points and returns corrected coordinates of these points, which can
be considered as samples of a smooth surface. The idea of the shape filter
developed in the present study is based on the representation of the mapping
of each bubble surface on a topologically equivalent object. We implemented
and tested filters for shapes of genus zero (topologically equivalent to a
sphere), and genus one (topologically equivalent to a torus). As the idea of
the filter is the same and just the basis functions are different, we
describe the method for the former case only (the spherical filter).

Consider a closed surface $S_{l}$ of the $l$th bubble, $l=1,...,M$, which is
topologically equivalent to a sphere. We denote a unit sphere as $S_{u},$
and spherical coordinates on this sphere as $\theta $ and $\varphi $. Thus,
the surface $S_{l}$ can be described parametrically as
\begin{eqnarray}\label{eq3_16}
&& \mathbf{r}=\mathbf{R}_{l}(\theta ,\varphi )=\left( X_{l}\left( \theta
,\varphi \right) ,Y_{l}\left( \theta ,\varphi \right) ,Z_{l}\left( \theta
,\varphi \right) \right) ,\\
&&  0\leqslant \theta \leqslant \pi ,\quad
0\leqslant \varphi <2\pi . \notag
\end{eqnarray}%
Since the surface is closed, function $\mathbf{R}_{l}$ is a periodic
function of $\varphi $ and obeys a spherical symmetry
\begin{eqnarray}\label{eq3_17}
&& \mathbf{R}_{l}(\theta ,\varphi )=\mathbf{R}_{l}(\theta ,\varphi +2\pi
),\\
&&  \mathbf{R}_{l}(\theta +\pi ,\varphi )=\mathbf{R}_{l}(\theta ,\varphi
+\pi ).   \notag
\end{eqnarray}%
Assuming that $\mathbf{R}_{l}\in L_{2}\left( S_{u}\right) $, we can expand
it into a series of spherical harmonics,%
\begin{eqnarray}\label{eq3_18}
&& X_{l}(\theta ,\varphi ) =\sum_{n=0}^{\infty
}\sum_{m=-n}^{n}X_{ln}^{m}Y_{n}^{m}\left( \theta ,\varphi \right) ,\quad
 \\
&& Y_{l}(\theta ,\varphi ) =\sum_{n=0}^{\infty
}\sum_{m=-n}^{n}Y_{ln}^{m}Y_{n}^{m}\left( \theta ,\varphi \right) ,  \notag
\\
&& Z_{l}(\theta ,\varphi ) =\sum_{n=0}^{\infty
}\sum_{m=-n}^{n}Z_{ln}^{m}Y_{n}^{m}\left( \theta ,\varphi \right) ,  \notag
\end{eqnarray}%
where $X_{ln}^{m},Y_{ln}^{m},$ and $Z_{ln}^{m}$ are the expansion
coefficients, and $Y_{n}^{m}\left( \theta ,\varphi \right) $ are the
spherical harmonics.

Consider now the shape filtering procedure. First, we should truncate the
infinite series (\ref{eq3_18}) by limiting the values of $n$ to the first $%
p_{f}$ modes, $n=0,...,p_{f}-1$. The truncation number $p_{f}$ can also be
called ``filter bandwidth''. Such finite series can be represented in the
form (we write this for the $x$ coordinate only, as the expressions for the
other coordinates are similar)

\begin{eqnarray}\label{spherical4}
&& X_{l}(\theta ,\varphi
)=\sum_{n=0}^{p_{f}-1}\sum_{m=-n}^{n}X_{ln}^{m}Y_{n}^{m}\left( \theta
,\varphi \right) \\
&& =\sum_{j=1}^{P_{f}}X_{lj}Y_{j}\left( \theta ,\varphi
\right) ,\quad P_{f}=p_{f}^{2},  \notag
\end{eqnarray}%
where the multiindex $j=(n+1)^{2}-n+m,$ $n=0,...,p_{f}-1,$ $m=-n,...,n$ is
used to map a pair of indices ($n,m$) to a single index.

Second, we note that if the surface at some moment of time (e.g., at $t=0$)
is initialized, then any point on the surface $S_{l}$ described as $\mathbf{r%
}_{li}(t)=\mathbf{R}_{l}(\theta _{li},\varphi _{li},t),$ $i=1,...,N_{d},$
evolves at constant $\theta _{i}$ and $\varphi _{i}$ specific for this
point. So the correspondence between the point index and the spherical
angles is established. We select now $p_{f}$ $<N_{d}^{1/2}$, to get
overdetermined systems for each Cartesian coordinate (only for $x$ is
displayed),
\begin{eqnarray}\label{spherical5}
&& \sum_{j=1}^{P_{f}}X_{lj}Y_{j}\left( \theta _{li},\varphi _{li}\right)
=x_{li},\\
&& i=1,...,N_{d},\quad l=1,...,M.\text{ }  \notag
\end{eqnarray}%
These equations also can be written in the matrix-vector form%
\begin{eqnarray}\label{eq3_22}
&&\mathbf{G}_{l}\mathbf{X}_{l} =\mathbf{x}_{l},\quad \mathbf{X}_{l}=\left(
\begin{array}{c}
X_{l1} \\
... \\
X_{lP_{f}}%
\end{array}%
\right) ,\quad \mathbf{x}_{l}=\left(
\begin{array}{c}
x_{l1} \\
... \\
x_{lN_{d}}%
\end{array}%
\right) ,\quad  \\
&& \mathbf{G}_{l} =\left(
\begin{array}{ccc}
Y_{1}\left( \theta _{l1},\varphi _{l1}\right) & ... & Y_{P_{f}}\left( \theta
_{l1},\varphi _{l1}\right) \\
... & ... & ... \\
Y_{1}\left( \theta _{lN_{d}},\varphi _{lN_{d}}\right) & ... &
Y_{P_{f}}\left( \theta _{lN_{d}},\varphi _{lN_{d}}\right)%
\end{array}%
\right) .\quad  \notag
\end{eqnarray}%
Third, we solve the overdetermined system using the least squares method,%
\begin{eqnarray}\label{eq3_23}
&& \mathbf{X}_{l}=\mathbf{G}_{l}^{(-1)}\mathbf{x}_{l},\quad \mathbf{G}%
_{l}^{(-1)}=(\mathbf{G}_{l}^{\ast }\mathbf{G}_{l}\mathbf{)}^{-1}\mathbf{G}%
_{l}^{\ast },
\end{eqnarray}%
where $\mathbf{G}_{l}^{(-1)}$ is the pseudoinverse and $\mathbf{G}_{l}^{\ast
}$ is the conjugate transpose of $\mathbf{G}_{l}$.

Finally, we compute the filtered values of the coordinates as%
\begin{eqnarray}\label{eq3_24}
&& \widehat{\mathbf{x}}_{l}=\mathbf{G}_{l}\mathbf{X}_{l}=\mathbf{G}_{l}\mathbf{G%
}_{l}^{(-1)}\mathbf{x}_{l}=\mathbf{F}_{l}\mathbf{x}_{l},\\
&& \widehat{\mathbf{y}}_{l}=\mathbf{F}_{l}\mathbf{y}_{l},\quad \widehat{\mathbf{%
z}}_{l}=\mathbf{F}_{l}\mathbf{z}_{l},  \notag
\end{eqnarray}%
where
\begin{eqnarray}\label{eq3_25}
&& \mathbf{F}_{l}=\mathbf{G}_{l}(\mathbf{G}_{l}^{\ast }\mathbf{G}_{l}\mathbf{)}%
^{-1}\mathbf{G}_{l}^{\ast },\quad l=1,...,M
\end{eqnarray}%
is the filtering matrix of size $N_{d}\times N_{d}$, or the filter of
bandwidth $p_{f}$.

Several remarks can be done here. First, the same filter can be applied to
any surface function provided by samples. For example, we have samples of
potential $\phi _{li}=\phi _{l}\left( \mathbf{r}_{li}\right) $ on the
surface of the $l$th bubble. Since the potential can also be expanded over
the spherical harmonics and the filter of the same bandwidth can be applied,
we have%
\begin{eqnarray}\label{spherical6}
&& \widehat{\mathbf{\phi }}_{l}=\mathbf{F}_{l}\mathbf{\phi }_{l},\quad \mathbf{%
\phi }_{l}=\left( \phi _{l1},...,\phi _{lN_{d}}\right) ^{T},
\end{eqnarray}%
where superscript $T$ denotes transposition.

Second, computation of the filter for the $l$th bubble has complexity $%
O\left( N_{d}^{3}\right) $. However, during the surface evolution in the
absence of any regridding $\theta _{i}$ and $\varphi _{i}$ for each point
are constant, so the filter should be computed only once, stored, and used
to smooth the surface functions at any moment of time (if the regridding is
needed the filter can be recomputed). Since the cost of a single
matrix-vector multiplication is $O\left( N_{d}^{2}\right) $, the
computational cost of filtering is $O\left( N_{d}^{2}M\right) $. This cost
can be compared with the cost of the FMM, which is formally $O\left(
N_{d}M\right) $, but it has a large asymptotic constant of the order of 54$s$
in optimal settings, where $s$ is the FMM clustering parameter \cite%
{Gumerov2008}. For example, at $N_{d}=600$ and $s=100$, the filtering cost
is about 10\% of the FMM cost. Since the iterative solution requires several
FMM calls per time step, the relative cost of filtering is really small. In
the present algorithm, we applied the filter twice, each time when the
right-hand side of Eqs (\ref{eq3_14}) and (\ref{eq3_15}) is called (first,
to filter input data $\mathbf{r}$ and $\phi $ and, second, to filter the
output data, i.e., the computed right-hand sides of these equations).

Third, when studying large bubble clouds, the initial bubble shapes can be
very similar (typically, all bubbles are spheres at $t=0$), or all bubbles
can be classified into several groups of bubbles having similar initial
shapes. In such cases, the actual number of filters needed reduces
dramatically. Indeed, only one shape filter is necessary, when all bubbles
have the same initial shape (the radius of the sphere or any length scaling
factor does not affect the filtering matrix), and only $K$ different shape
filters are needed if there are $K$ different initial shapes of the bubbles.
So, the total cost of computing the filtering matrices for all bubbles is $%
O\left( N_{d}^{3}K\right) $, which is much smaller than $O\left(
N_{d}^{3}M\right) $ at $K\ll M$.

Fourth, the toroidal filter is designed in the same way, but instead of the
spherical transform the 2D Fourier transform is used (here we have $2\pi $%
-periodic functions of angles $\varphi $ and $\theta $ describing positions
of the points on a unit torus). So, for the filter of bandwidth $p_{f}$ we
have%
\begin{eqnarray}\label{spherical7}
&& X_{l}(\theta ,\varphi
)=\sum_{n=-p_{f}/2}^{p_{f}/2}\sum_{m=-p_{f}/2}^{p_{f}/2}X_{ln}^{m}F_{n}^{m}%
\left( \theta ,\varphi \right) ,\text{\quad }\\
&& F_{n}^{m}\left( \theta
,\varphi \right) =e^{i(n\theta +m\varphi )}.  \notag
\end{eqnarray}%
Steps described by Eq.~(\ref{spherical4}) and below then can be repeated
with slight modifications.

Fifth, for real functions the real spherical harmonic basis can be used.
Similarly, for the toroidal filter, one can use the real trigonometric
basis. But it is also noticeable, that in any case the use of the real or
complex basis practically does not affect the overall algorithm complexity
since the filtering matrices $\mathbf{F}_{l}$ anyway are real and symmetric,
and as soon as they precomputed and stored the way how they are obtained
does not matter.

Finally, parameter $p_{f}$ should be selected reasonably small to provide
good smoothing (substantial oversampling), but it also should be moderately
high to enable tracking of essential shape variation and reduce the memory
(no excessive oversampling), e.g., $p_{f}\lesssim \left( N_{d}/2\right)
^{1/2}$. Note also that if $p_{f}^{2}=N_{d}$ then the filtering matrix is
just the identity matrix (no filtering).

\subsection{Iterative solver}

Equations~(\ref{eq3_2}) can be solved using different iterative methods.
Krylov methods require computation of the matrix-vector product $\mathbf{LX}$%
, where $\mathbf{X}$ is some input vector, and $\mathbf{L}$ is the system
matrix. In the present study, we tried the unpreconditioned general minimal
residual method (GMRES) \cite{Saad1986} and preconditioned flexible GMRES %
\cite{Saad1993}. For the cases reported in the present paper, the former
method converges in a few iterations when the initial guess is provided by
the solution at the previous time step.

\subsection{Fast multipole method}

In the conventional BEM, matrices $\mathbf{L}$ and $\mathbf{M}$ (Eq.~(\ref%
{eq3_2})) should be computed and saved to solve the linear system either
directly or iteratively. The memory needed to store these matrices is fixed
and is not affected by the accuracy imposed on the computation of the
surface integrals, which should be computed only once for a given mesh. The
memory limits impose severe constraints on the size of the computable
problems, so the large-scale problems can be solved iteratively by methods
where the matrix-vector product is computed ``on the fly'' without matrix
storage.

The matrix-vector products (MVPs) involving the BEM matrices can be computed
using the FMM. Further, in the context of the FMM, we consider a single
matrix
\begin{eqnarray}
&& \mathbf{A}=f\mathbf{L}+g\mathbf{M,}  \label{eq3_27}
\end{eqnarray}%
which turns to $\mathbf{L}$ and $\mathbf{M}$ at $f=1,\;g=0$, and $f=0,\;g=1$ and
MVP $\mathbf{AX}$, where $\mathbf{X}$ is some input vector. There exist
several approaches to the use of the FMM in the BEM. Traditional methods use
factorization of the BEM integrals \cite{Nishimura2002}, which requires some
modifications of a ``standard'' FMM designed for summation of monopoles and
dipoles. Such standard FMM codes currently are available in the form of open
source or commercial software, which can be considered as black box FMM
solvers. Low order approximations of non-singular integrals (see Eq. (\ref%
{ns1})) can use black box FMMs without any modifications. In a recent paper %
\cite{Adelman2017} an algorithm using such black box FMM solvers, but for
arbitrary order approximation of non-singular BEM integrals is developed.
The method introduces ``correction matrices'', which are the difference of
the BEM matrices computed using high and low order quadratures. It is shown
that such correction matrices are sparse, and the algorithm can be used for
large scale simulations.

In the present study, we use only standard FMMs for the Laplace equation in
three dimensions (with or without GPUs), which detailed description can be
found elsewhere (e.g., \cite{Greengard1987,GumerovL2005}). More precisely,
the FMM used in this study is implemented as described in \cite{Gumerov2008}%
, where a part of the algorithm was accelerated using a GPU while the other
part of the algorithm was accelerated using Open MP for a multicore CPU.

Briefly, the FMM can be described as follows. The entries of matrix $\mathbf{%
A}$ ($A_{ij}$) in Eq. (\ref{eq3_27}) can be treated as some interaction
coefficients between the $j$th and the $i$th collocation points, which we
call ``sources'' and \ ``receivers'', respectively. The dense matrix $%
\mathbf{A}$ can be formally represented as $\mathbf{A=}$ $\mathbf{A}%
^{(dense)}+$ $\mathbf{A}^{(sparse)}$, where $\mathbf{A}^{(sparse)}$ accounts
for interactions between the receivers and the sources located in some
neighborhood of these receivers, while $\mathbf{A}^{(dense)}$ accounts for
the rest of interactions. Respectively, in an iterative solver computation
of MVP $\mathbf{AX}$ can be split into computations of $\mathbf{A}^{(dense)}%
\mathbf{X}$ and $\mathbf{A}^{(sparse)}\mathbf{X}$. Both products can be
computed at $O(N)$ computational cost, which is obvious for $\mathbf{A}%
^{(sparse)}\mathbf{X}$. Computation of $\mathbf{A}^{(dense)}\mathbf{X}$ is
less trivial, as it requires partitioning of the computational domain with a
hierarchical data structure (boxes), multipole expansions of the fields
generated by the sources, translations, and evaluation of the local
expansions. The multipole and local basis functions for the Laplace equation
are proportional to the spherical harmonics of degree $n=0,1,...$ and order $%
m=-n,...,n$ (similarly to Eq.~(\ref{eq3_18})). The infinite expansions are
truncated to the first $p_{fmm}$ degrees ($n=0,...,p_{fmm}-1$), where $%
p_{fmm}$ is the truncation number. Such truncation enables operation with
relatively compact representations of functions and, as the total number of
expansion and translation operations is $O(N)$, results in $O(N)$ algorithm
complexity. Of course, the truncation of the infinite series introduces
errors, which are controlled by $p_{fmm}$. This value also affects the
asymptotic constant in the algorithm complexity, and a reasonable balance
between the accuracy and speed can be found via optimization.

The FMM uses a data structure. The computational domain is scaled to a unit
cube (level $l=0$) and recursively subdivided using the octree structure to
level $l_{\max }$. Level $l$ contains $8^{l}$ boxes. The source and receiver
data structures exclude empty boxes, allow a fast neighbor finding, and
provide interaction lists (e.g., via bit-interleaving \cite{Gumerov2005}
followed by a sorting algorithm). For a fixed mesh (or matrix $\mathbf{A}$),
this part of the algorithm should be called only once (in contrast to
computation of MVP $\mathbf{AX}$ at different $\mathbf{X}$ in the iterative
solver).

\subsection{Parallelization}

There is a substantial difference between the parallelization of the FMM on
computing systems with shared memory and distributed memory. The distributed
memory systems are typical for clusters consisting of many computing nodes
communicating via the MPI. The communication overheads here can be
substantial. Moreover, the algorithm should be carefully designed to provide
more or less even loads for the nodes. Most studies related to
parallelization of the FMM are about distributed memory systems, and they
address issues of efficient load balancing and communications.

In a shared memory system, all processes have relatively fast access to the
global system memory, which contains information about the entire data
structure and makes data computed by each process available to all processes
almost immediately. A typical example of such a system is a multicore CPU.
The parallel algorithms for them can be much more straightforward (e.g., use
parallelization of the loops of a serial algorithm using the OMP). We tested
such schemes and found that almost all loops of the serial FMM can be
parallelized in this way to achieve high parallelization efficiency.
Modification of the serial algorithm is needed only for sorting algorithms
used for generation of the data structure. However, for the BEM this is not
critical as the cost of sorting is relatively low and it is amortized over
several iterations within a time step.

In this context, the efficiency of use of GPUs should be reconsidered.
Indeed, in study \cite{Gumerov2008} it was shown that 30-60x accelerations
of the FMM compared to a serial algorithm can be obtained using a single
GPU. However, these days CPUs with, say eight cores are typical and 32 or 64
core machines also available to the researchers. Despite the GPU performance
also increased compared to the year 2007 the relative efficiency of the GPU
parallelization is substantially lower compared to multicore CPUs. Of
course, there are always some solutions (usually costly) with many GPUs in
one workstation, where the ratio of the CPU cores and GPUs should be a
criterion for the efficiency of graphics processors in the FMM.

Profiling of the GPU efficiency for the different parts of the FMM\ \cite%
{Gumerov2008} shows that a significant acceleration can be obtained when
using GPU for computation of $\mathbf{A}^{(sparse)}\mathbf{X}$, which is due
to both the actual acceleration, and the reduction of the depth of the
octree $l_{\max }$. The reported 2-10 times acceleration of the translation
operations on the GPU compared to a single core CPU can be easily achieved
on a multicore CPU. By this reason, in the present implementation, we used
GPU only for the part where it is the most efficient, namely just for the
sparse matrix-vector product, while the other parts of the algorithm were
accelerated using the OMP.

\subsection{Performance of MVP accelerators}

Some tests of the FMM implemented on CPU/GPU and parallelized on the CPU via
OMP are discussed below. The times are measured on a workstation equipped
with Intel Xeon 5660 2.8 GHz CPU (12 physical cores), 12 GB RAM, and one GPU
NVIDIA Tesla K20 (5 GB of global memory). All GPU computations are conducted
with single and double precision. In all cases reported in this section, we
used monopole sources distributed randomly and uniformly inside a cubic
domain.

\begin{table}[tbp]
\caption{The relative error of the FMM}
\label{Table1}
\begin{center}
\begin{tabular}{|p{1cm}|p{1cm}|p{1cm}|p{1cm}|p{1cm}|}
\hline
\hline
\multirow{2}{*}{$p_{fmm}$} & \multicolumn{2}{|l|}{$N=131,072$} & \multicolumn{2}{|l|}{$N=1,048,576$} \\ \cline{2-5}
& single & double  &
single  & double  \\ \hline\hline
4 & $5\cdot 10^{-4}$ & $7\cdot 10^{-4}$ & $7\cdot 10^{-4}$ & $7\cdot 10^{-4}$ \\
8 & $8\cdot10^{-6}$  & $10^{-5}$ & $8\cdot 10^{-6}$ & $10^{-5}$ \\
12 & $4\cdot10^{-7}$  & $4\cdot 10^{-7}$ & $5\cdot 10^{-7}$ & $5\cdot 10^{-7}$ \\
16 & $2\cdot10^{-7}$  & $3\cdot 10^{-8}$ & $6\cdot 10^{-8}$ & $3\cdot 10^{-8}$ \\
20 & $2\cdot10^{-7}$  & $4\cdot 10^{-9}$ & $5\cdot 10^{-8}$ & $4\cdot 10^{-9}$ \\ \hline\hline
\end{tabular}%
\end{center}
\end{table}

The FMM trades the accuracy for speed. The runtime of the FMM depends on the
truncation number $p_{fmm}$ and also on the precision of calculations of \\$%
\mathbf{A}^{(sparse)}\mathbf{X}$ performed on the GPU. A general rule for
faster computations is to use $p_{fmm}$ as small as possible and single
precision on GPU if possible. However, both of these parameters affect the
accuracy of the result, which in any case should be the first thing to
consider. Table \ref{Table1} shows the relative $L_{2}$-norm error of the
MVP as a function of these parameters. The accuracy of $N$-point computing
is estimated using the direct evaluation of the product at $\sqrt{N}$
checkpoints as a reference. In this case, the estimated error does not
depend substantially on $N$. The table also shows that up to values $%
p_{fmm}=12$ the accuracy of the MVP practically is not affected by the
precision of GPU computing. Modern GPUs usually perform single precision
computations about two times faster compared to double precision
computations. Hence, in an ideally balanced FMM ($l_{\max }$ is selected to
provide the same costs of $\mathbf{A}^{(dense)}\mathbf{X}$ and $\mathbf{A}%
^{(sparse)}\mathbf{X}$), use of single precision computing for $\mathbf{A}%
^{(sparse)}\mathbf{X}$ should accelerate the overall algorithm approximately
1.5 times. Such a balanced algorithm is hardly achievable in practice since $%
l_{\max }$ changes discretely. Also, the distribution of the sources, the
size of blocks of threads in GPU, etc., affect the actual accelerations.

\begin{figure}[h]
\begin{center}
\includegraphics [scale=1]{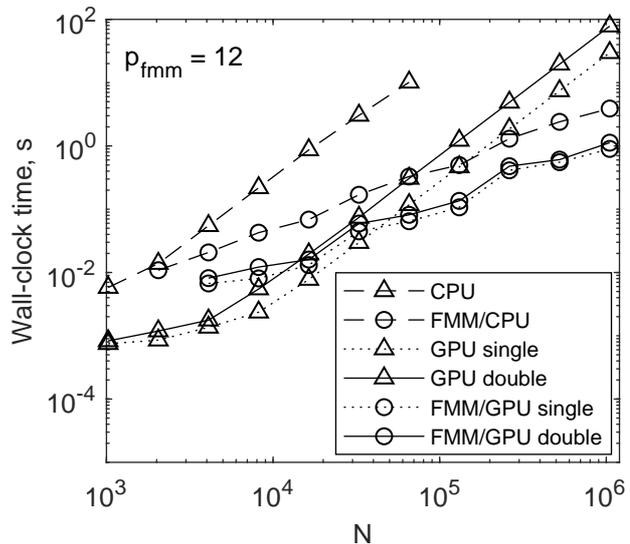}
\end{center}
\caption{The wall-clock time required for a single MVP using different
hardware and algorithmic accelerators (the time for generation of the data
structure is not included). The data points are distributed randomly inside
a unit cube. The time is measured on a PC equipped with a 12-core CPU (Intel
Xeon 5660 2.8 GHz) and one GPU (NVIDIA Tesla K20).}
\label{Fig1}
\end{figure}

\begin{table}[tbp]
\caption{Acceleration of the MVP using different methods ($p_{fmm}=12$)}
\label{Table2}
\begin{center}
\begin{tabular}{|p{1.2cm}|p{0.8cm}|p{0.8cm}|p{0.8cm}|p{0.8cm}|p{0.8cm}|}
\hline
$N$ & \multicolumn{2}{|l|}{GPU} & FMM &
\multicolumn{2}{|l|}{FMM+GPU} \\ \hline\hline
& single & double &  & single & double \\
4,096 & 40 & 31 & 3 & 8 & 7 \\
16,384 & 111 & 44 & 13 & 66 & 54 \\
65,536 & 86 & 33 & 31 & 157 & 125 \\
262,144 & 87 & 33 & 125 & 395 & 339 \\
1,048,576 & 87 & 33 & 667 & 2896 & 2309 \\ \hline\hline
\end{tabular}%
\end{center}
\end{table}

Three methods are considered to accelerate the MVP: first, brute-force
hardware (GPU) acceleration; second, algorithmic (FMM) acceleration (the
time for generation of the data structure is neglected); and, third,
algorithmic (FMM) + hardware (GPU) acceleration. Figure~\ref{Fig1}
demonstrates the dependence of the runtime for the MVP using these three
approaches (also the times for single and double precision GPU computing are
measured separately). It is seen that the complexity of the brute-force MVP
implemented on the CPU or GPU is quadratic in $N$ while it is close to
linear for the FMM implemented on CPU/GPU. In all cases the FMM was
optimized (the optimal $l_{\max }$ was determined for each $N$; it is
different for CPU and GPU, see \cite{Gumerov2008}). Table \ref{Table2} shows
the accelerations achieved using the three methods mentioned. Here a CPU
code implementing brute-force MVP is taken as a reference (for large $N$ the
data for the reference case are extrapolated proportionally to $N^{2}$). It
is seen that at large enough $N$ the GPU acceleration of the brute-force
method stabilizes near some constants, and single precision computing 2.6
times faster than the double precision. The use of single precision in the
GPU accelerated FMM brings smaller gains in speed. This table along with
Fig. \ref{Fig1} also shows that for problem sizes $N\lesssim 30,000$ the use
of a such complex algorithm as the FMM is not justified since the
brute-force use of a GPU delivers the same or better accelerations of the
reference code.

\section{Numerical examples}

The method described above is validated in many tests, some of which are
presented below and can be used as benchmark cases when comparing different
bubble dynamics codes. All numerical results presented in this paper (except
the comparisons with the benchmark cases of Bui et al \cite{Bui2006}) are
obtained for air bubbles in water ($\rho =1000$~kg/m$^{3}$, $\sigma =0.073$%
~Pa/m, $\kappa =1.4$) under atmospheric pressure ($p_{0}=10^{5}$~Pa) and
zero gravity. To illustrate computations, the dimensionless coordinates $%
x^{\prime }=x/a_{0},\;y^{\prime }=y/a_{0},\;z^{\prime }=z/a_{0}$ are used,
where $a_{0}$ is the initial bubble size/radius.

\subsection{Single spherical bubble in an acoustic field}

\begin{figure}[t]
\begin{center}
\includegraphics [scale=1]{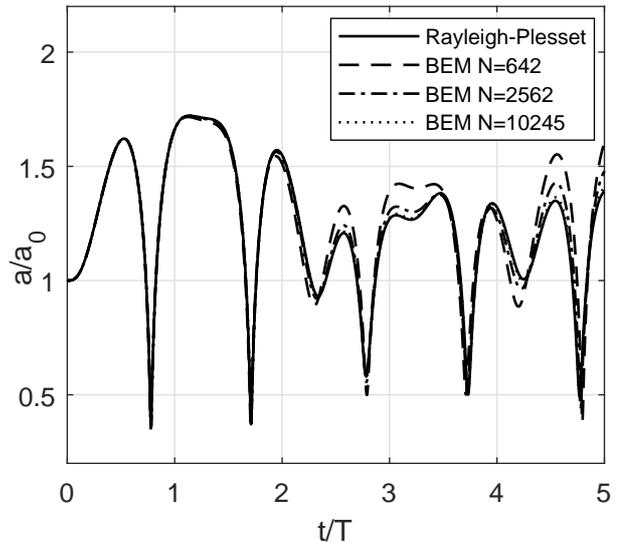}
\end{center}
\caption{A comparison of the numerical solution at different surface
discretizations with the solution of the Rayleigh-Plesset equation for a
spherical bubble.}
\label{Fig2}
\end{figure}

As a starting point, some tests for a single spherical bubble ($a_{0}=10$ $%
\mu m$) under the action of an acoustic field are conducted. The obtained
results are compared with the solution of the Rayleigh-Plesset equation \cite%
{Plesset1977} at zero viscosity,
\begin{eqnarray} \label{eq4_1}
&& a\ddot{a}+\frac{3}{2}\dot{a}^{2}=\left[ p_{g0}\left( \frac{%
a_{0}}{a}\right) ^{3\kappa }-p_{\infty }\left( t\right) -\frac{2\sigma }{a}%
\right] , \\
&& a(0)=a_{0},\quad \dot{a}(0)=0. \notag
\end{eqnarray}%
Figure~\ref{Fig2} compares the dynamics of bubble radius using different
number of boundary elements and the ``exact'' solution for the amplitude of
the acoustic field $P_{a}=p_{0}$ and frequency $200$ kHz (period $T=5$ $\mu s
$) (``exact'' means error controlled numerical solution of ODE (\ref{eq4_1}%
)). It is seen that the discretization has a little effect on the computed
results at $t\leq 2T$, while for $t>2T$ substantially high $N$ is needed to
reproduce the spherical bubble dynamics accurately. This phenomenon can be
explained by the fact that the bubble shape approximated by the mesh is not
exactly spherical. So, there exists some energy transfer between the volume
and shape modes, which manifests itself at smaller times for smaller $N$.

It is noticeable that even this simple case cannot be computed without
surface smoothing techniques. For the range of $N$ used in this example, the
mesh was destabilized within 40-50 time steps without shape filtering (52
time steps at $N=642$, 40 time steps at $N=2562$, and 37 time steps at $%
N=10245$). The stability condition (\ref{eq3_12}) requires smaller time
steps at higher surface discretization. So the time $t_{\max }$ for
computable solution without surface smoothing decreases at increasing $N$ ($%
t_{\max }/T=0.132,$ $0.0512,$ and $0.0229$ at $N=642,$ $2562,$ and $10245$,
respectively). Utilization of the spherical filter at each time step with
reasonable $p_{f}$ removes this instability and computations can\ proceed to
a user-specified $t_{\max }.$

The term ``reasonable $p_{f}$'' requires some elaboration. In this and other
cases reported below we found that at large enough $p_{f}$ ($p_{f}>10$) the
surface destabilizes anyway (at later times compared to the computations
without the surface smoothing, since the lower frequency shape instabilities
develop anyway), while at small enough $p_{f}$ ($p_{f}<4$) details of shape
deformation can be lost or reproduced incorrectly. We varied parameters $%
N_{d}$ and $p_{f}$ to check the correctness of the results. For example,
computations of formation of jets in interacting bubbles are consistent for
range $p_{f}=4,...,9$ and $N_{d}=642$ and $2562$. In the cases reproted
below, one should assume $p_{f}=6$ and $N_{d}=642$ (if not stated otherwise).

\subsection{Comparisons with some reported cases}

There exist many studies of bubble dynamics in three dimensions
demonstrating various shapes of the bubbles. Bui et al \cite{Bui2006} well
documented 7 cases of different bubble configurations, which can be
considered as benchmark cases. We computed several of these cases (only two
of them are reported below) and found satisfactory qualitative and
quantitative agreement with the data reported in \cite{Bui2006}. All cases %
\cite{Bui2006} are computed for simplified bubble dynamics model (underwater
explosion resulting in a relatively large bubble), which follows from Eqs (%
\ref{eq2_8})-(\ref{eq2_10}) at zero gravity, zero surface tension, and $%
F(t)=const.$ These equations coincide with the dimensionless equations of
Bui et al at $F(t)\equiv 1$, $\epsilon =p_{0}/\rho $, $\lambda =\kappa $,
where $\epsilon $ and $\lambda $ are the strength parameter and the gas
polytropic exponent. All benchmark cases computed at $\epsilon =97.52$ and $%
\lambda =1.25$. Hence, if in our computations we set $\rho =10^{3}$ kg/m$%
^{3} $, $p_{0}=0.9752\cdot 10^{5}$ Pa, then all dimensional variables in
basic SI units (lengths in meters, time in seconds), will be the same as the
respective dimensionless variables of the cited work.

\begin{figure}[t]
\begin{center}
\includegraphics [scale=0.8]{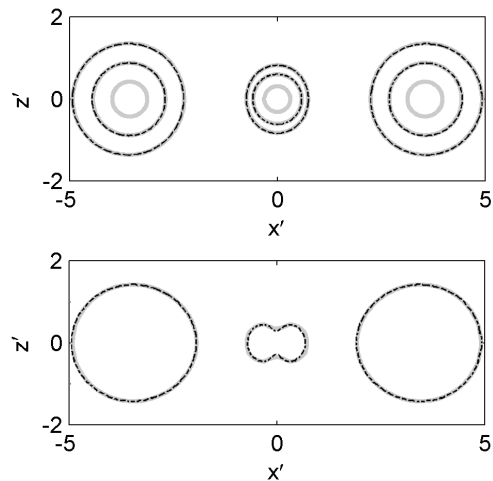}
\end{center}
\caption{A comparison of the solution for three bubbles (the gray lines)
with the Case 2 solution of Bui et al (2006) (the dashed lines) at the
moments of time $t=0.05$~$s$ and $t=0.758$~$s$ (on the left plot) and at $%
t=2.65$~$s$ (on the right plot). The contour at $t=0.05$ $s$ of Bui et al
corresponds to the shape at $t=0.3$~$s$ in the present computations (for
comparisons, the shape at $t=0.05$~$s$ is also shown).}
\label{Fig3}
\end{figure}

Figure \ref{Fig3} compares the results for the benchmark Case 2 (three
equispaced bubbles in a row of initial distance between the centers 3.55~$m$%
, and radii for the central bubble $0.1499$~$m$ and $0.2379$~$m$
for the outer bubbles. It is seen that the shapes and positions of the
bubbles agree well. The difference is explainable by different
discretizations of the surface and the different numerical techniques
(different ways of evaluation of boundary integrals, different surface
smoothing procedures, etc.). Our estimates show that these factors can
explain the relative differences in computed results of the order of several
percents. It is also noticeable that the relative errors of the BEM alone
(just because of the surface discretization by flat triangles) is a few
percents, which also agrees with the estimates of \cite{Bui2006}. Note a
misprint in \cite{Bui2006} for this case, as two moments of time $t=0.05$~$s$
and $0.758$~$s$ are reported for the growth phase. In fact, the reported
contour at $0.05$~$s$ corresponds to a shape at $0.3$~$s$ in our
computations (for comparisons we also show our result at $0.05$~$s$).

\begin{figure}[t]
\begin{center}
\includegraphics [scale=0.8]{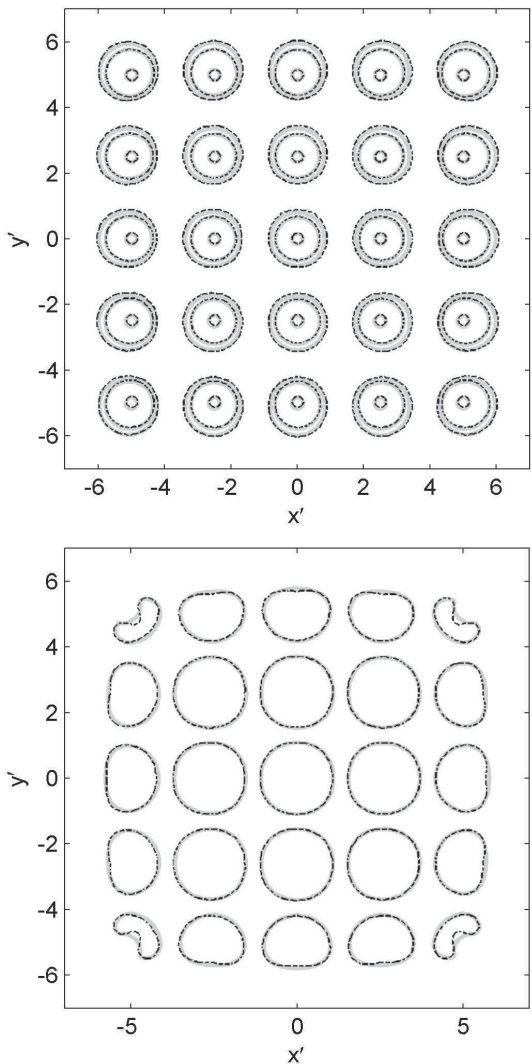}
\end{center}
\caption{A comparison of the solution for 25 bubbles (the gray lines) with
the Case 7 solution of Bui et al (2006) (the dashed lines) at the moments of
time $t=0.01$~$s,$ $t=0.473$~$s$, and $t=0.923$~$s$ (on the left plot) and $%
t=2.854$~$s$ (the present work) and at $t=2.935$~$s$ (Bui et al) (on the
right plot).}
\label{Fig4}
\end{figure}

Comparisons of the present computations and the results reported in \cite%
{Bui2006} for the benchmark Case 7 ($25$ bubbles of initial equal size $%
0.1499$~$m$ at $t=0$ are arranged into a $5\times 5$ 2D grid with the
minimal distance 2.5 $m$ between the bubble centers) are shown in Fig. \ref%
{Fig4}. Again, one can see a satisfactory agreement between the results.
Note that the pictures for the collapse stage obtained at time $t=2.854$~$s$%
, while \cite{Bui2006} reported the nondimensional time $t^{\prime }=2.939$.
The difference between these moments is less than 3\% and can be explained
by the BEM errors and the differences in algorithms. The reason why the
present computations were not continued beyond $t=2.854$~$s$ is that the
corner bubbles changed their topology between $t=2.854$~$s$ and $t=2.939$~$s.
$

\subsection{Three bubbles in an acoustic field}

\begin{figure}[t!]
\begin{center}
\includegraphics [scale=1]{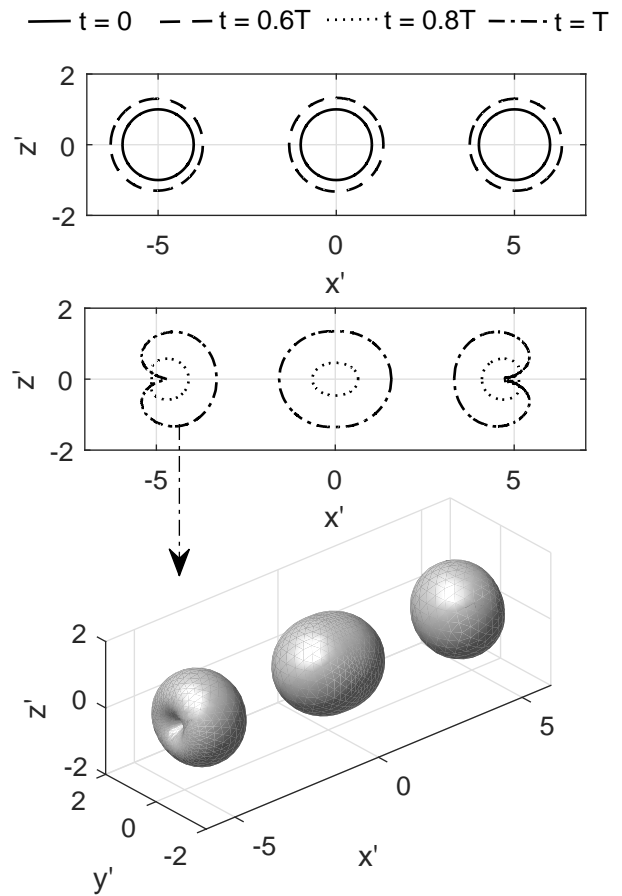}
\end{center}
\caption{On the upper plots: the projections of the bubble shapes in an
acoustic field at different moments of time $t$ for the test case a) ($T$ is
the period of the acoustic field). The bottom plot showing shapes in
3D corresponds to $t=T$. }
\label{Fig5}
\end{figure}

\begin{figure}[t!]
\begin{center}
\includegraphics [scale=1]{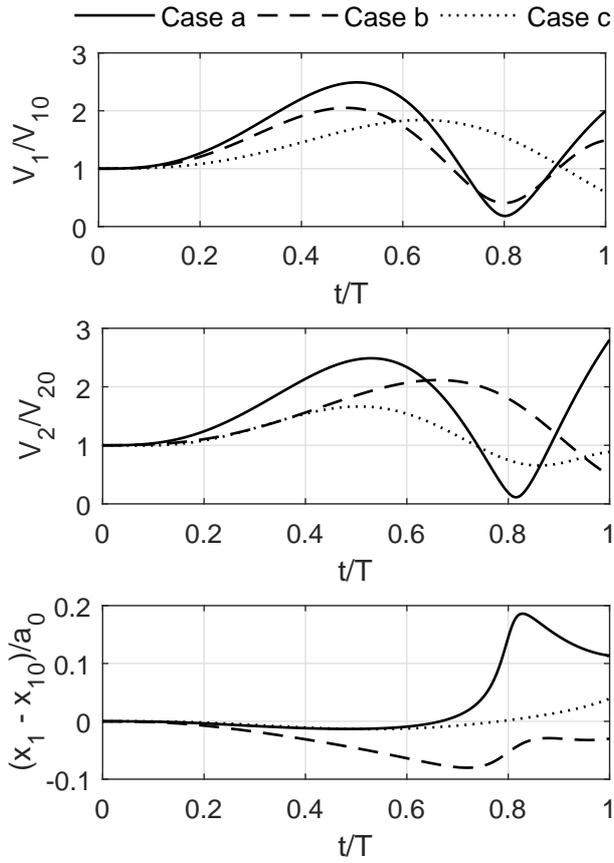}
\end{center}
\caption{The dynamics of the volume of bubble \#1 (on the top plot) and
bubble \#2 (in the middle plot), and the relative displacement of the mass
center of bubble \#1 (at the bottom plot) for test cases a) b) and c) (three
bubbles in an acoustic field of period $T$). }
\label{Fig6}
\end{figure}

To study bubble deformation and migration in an acoustic field we conducted
small-scale tests for three bubbles located on the $x$-axis, and labeled as
1,2,3 (the central bubble is labeled as ``\#2'' and the bubble with positive
$x$-coordinate of the center is labeled as ``\#3''). In all cases, the
distance $d$ between bubbles \#1 and \#2 is the same as the distance between
bubbles \#$2$ and \#$3$. Three cases are reported below: a) the initial
radii of all bubbles are the same, $a_{10}=a_{20}=a_{30}=a_{0}=10$~$\mu m$,
the initial distance between the bubble centers is $d=5a_{0}$; b) $%
a_{10}=a_{30}=a_{0}=10$~$\mu m$, $a_{20}=2a_{0}$, $d=6a_{0}$; c) $%
a_{20}=a_{0}=10$~$\mu m$, $a_{10}=a_{30}=2a_{0}$, $d=6a_{0}$. In all cases,
the acoustic field of frequency 200~$kHz$ ($T=5$~$\mu s$) has amplitude $%
P_{a}=0.7p_{0}$.

In case a) the qualitative picture is the following. At $0<t/T<0.55$ all
bubbles grow in volume and bubbles \#1 and \#3 slightly repel from the
central bubble while remaining spherical. At time $0.55<t/T<0.8$ bubbles
collapse and bubbles \#1 and \#3 move towards the central bubble, which
takes some elongated, ellipsoid-like shape. The bubble volume achieves its
minimum at $t/T<0.8$ after which all bubbles start to grow again, while
still attracting. In fact, this attraction of the bubble mass centers is due
to the formation of jets in bubbles \#1 and \#3 directed towards the central
bubble. Bubble \#$2$ is also non-spherical and has some prolate spheroid
shape. Contours of bubble shape projection on the $y=0$ plane and the shapes
of the bubbles at $t/T=1$ are shown in Fig. \ref{Fig5}.

The bubble dynamics in case b) is different from that observed in case~a).
Here the more massive bubble in the center remains almost spherical all the
time. The deformation of bubbles \#1 and \#3 is not so strong as in case~a).
Moreover, the jets are not formed in these bubbles, but at $t/T=1$ they
become pear-shaped tapering in the direction away from bubble~\#2. Case c)
is different from both cases a) and b). Here bubbles \#1 and \#3 are almost
spherical at $0\leqslant t/T\leqslant 1$, while qualitatively (not
quantitatively) the dynamics of bubble \#2 is similar to the dynamics of
bubble~\#2 in case~a). At $t/T=1$ it looks like a prolate spheroid. Figure~\ref{Fig6} shows the relative bubble volume as a function of time for all
three cases. It also demonstrates the dynamics of the mass center of bubble~\#1.

\subsection{Bubble clusters in an acoustic field}

\begin{figure}[t]
\begin{center}
\includegraphics [scale=1]{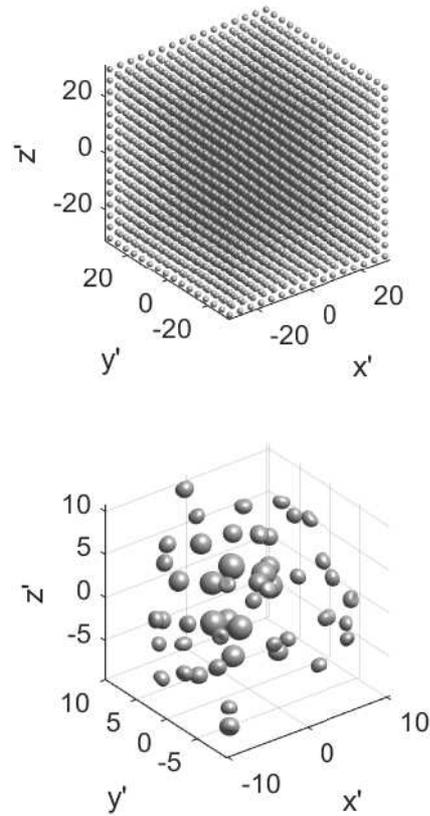}
\end{center}
\caption{Views of the largest $16\times 16\times 16$ regular cluster (on the
top plot) used in the present study at $t=0$. A view of a random ($M=51$) monodisperse
cluster in an acoustic field of period $T$ at $t=T$ (on the bottom plot). }
\label{Fig7}
\end{figure}

Using the present algorithm, we studied dynamics of monodisperse bubble clusters of
different sizes and configuration. Figure \ref{Fig7} shows the largest regular cluster used in the present tests ($M=4096$%
, $N=2,629,632$, $a_{0}=10$ $\mu m$), and a random monodisperse cluster ($M=51,$ $N=32,742,$ $%
a_{0}=10$ $\mu m$). Mention that for large clusters in acoustic fields
condition $D_{cl}/\lambda \ll 1$ ($D_{cl}$ is the cluster size and $\lambda $
is the acoustic wavelength) should hold to justify the assumption that the
liquid is incompressible. For example, the wavelength of 200 kHz sound in
water is $\lambda =$7.5 mm and for the largest case shown in Fig.~\ref{Fig7} we
have $D_{cl}/\lambda =0.08$ ($M=4096$).

\begin{figure}[t!]
\begin{center}
\includegraphics [scale=1]{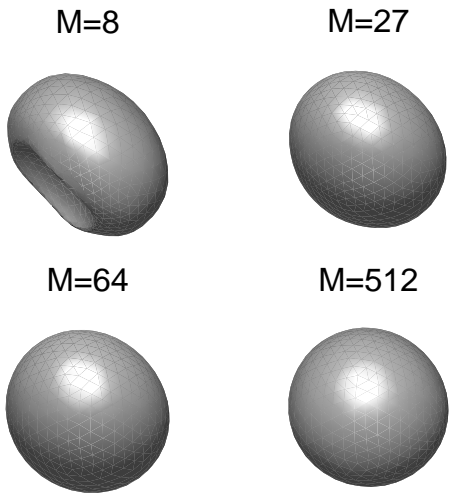}
\end{center}
\caption{The shapes of the corner bubbles in $n_{b}\times n_{b}\times n_{b}$
clusters ($n_{b}=2,3,4,8;$ $M=n_{b}^{3}$) at $t=0.9T$ ($M=8$) and $t=T$ ($%
M=27,64,512$)$.$}
\label{Fig8}
\end{figure}

\begin{figure}[t!]
\begin{center}
\includegraphics [scale=1]{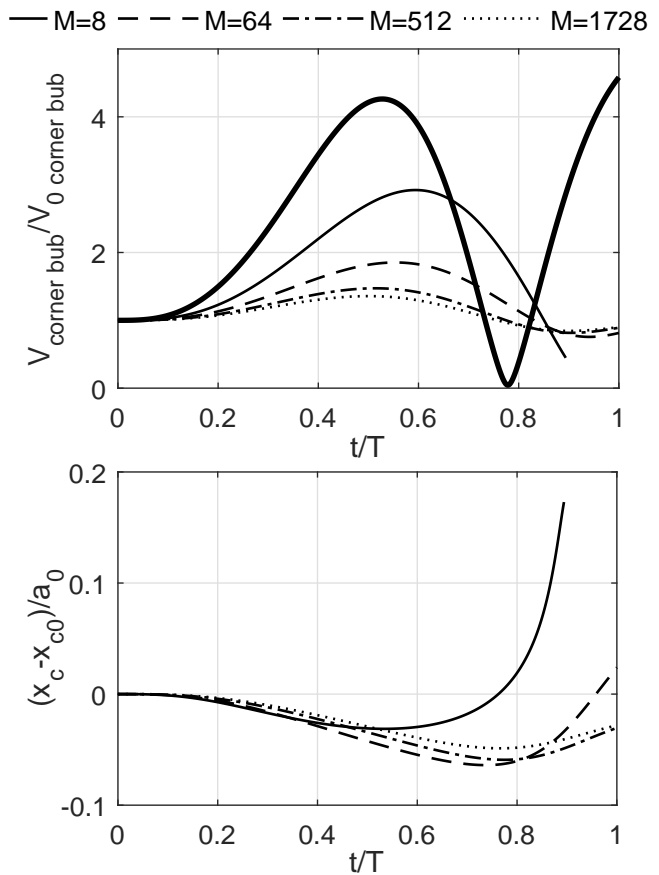}
\end{center}
\caption{ On the top plot: the dynamics of the volume of the corner bubble
in $n_{b}\times n_{b}\times n_{b}$ clusters ($n_{b}=1,2,4,8,12;$ $%
M=n_{b}^{3} $) in an acoustic field of period $T$. At the bottom plot: the $x
$-coordinate of the relative displacement of the mass center of the corner
bubble in the respective clusters. }
\label{Fig9}
\end{figure}

\begin{figure}[t!]
\begin{center}
\includegraphics [scale=1]{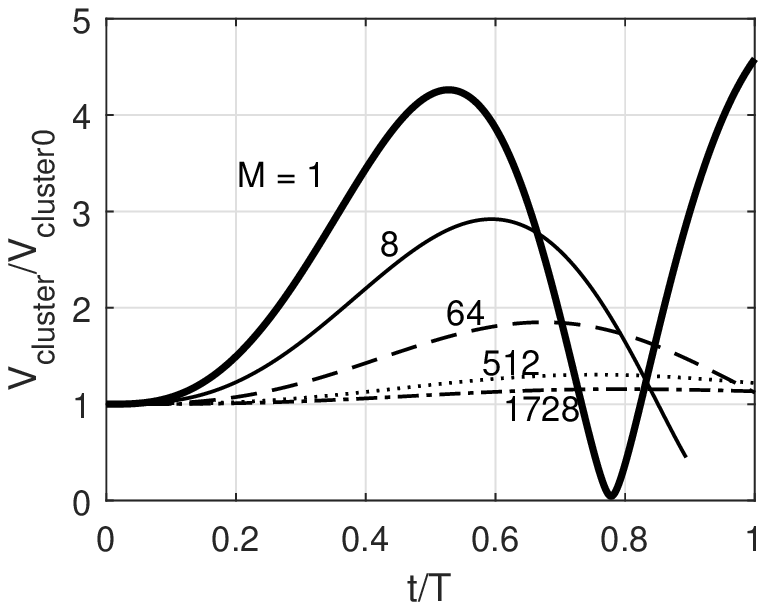}
\end{center}
\caption{The dynamics of the volume of $n_{b}\times n_{b}\times n_{b}$
bubble clusters ($n_{b}=1,2,4,8,12;$ $M=n_{b}^{3}$) in an acoustic field of
period $T$. }
\label{Fig10}
\end{figure}

Figures \ref{Fig8} - \ref{Fig10} provide some data, which can be used for
validation of this and other bubble dynamic codes. In these case bubble
cluster arranged in a $n_{b}\times n_{b}\times n_{b}$ cubic grid ($%
n_{b}=1,...,16$) consisting of spherical bubbles of initial size $a_{0}=10$ $%
\mu m$ is placed in an acoustic field of frequency 200~$kHz$ ($T=5$~$\mu s$)
and amplitude $P_{a}=p_{0}$. The initial distance between the closest bubble
centers ($n_{b}>1$) is $d=4a_{0}$. Figure \ref{Fig8} shows the shapes of the
corner bubble (the bubble with the minimum $x,y,z$ coordinates of the
centers) at $t/T=1$ ($t/T=0.9$ in case $n_{b}=2$, which is approximately the
moment of time when the bubble topology changed from spherical to toroidal). It is
seen that as the cluster size increases the shape of this bubble becomes
closer to spherical. In fact, the corner bubbles are the most deformed
bubbles in the cluster, so the other bubbles are more spherical. Figure \ref%
{Fig9} illustrates the dynamics of the volume of the corner bubble and the
change of the relative $x$-coordinate of its center. Hence, the larger
cluster, the weaker the response to variations of the external pressure
field. Figure \ref{Fig10} showing the dynamics of the volume of the entire
cluster also supports this conclusion.

\subsection{Performance}

Finally, we report some figures about the overall performance of the
developed code. The wall-clock time is measured for the work station
described above. The most critical issue here is the scaling of the code
with $N$. For this study one can use the cases with bubbles arranged in a $%
n_{b}\times n_{b}\times n_{b}$ cubic grid at different $n_{b}$.

Table \ref{Table3} shows profiling of the FMM/GPU code for $n_{b}=12$ and $%
n_{b}=16$ for two different accuracy settings. The faster option corresponds
to single precision GPU computing, $p_{fmm}=8$, and the tolerance for the
GMRES convergence $\epsilon _{gmres}=10^{-4}$. The other option is realized
using double precision GPU computing, $p_{fmm}=12,$ and $\epsilon
_{gmres}=10^{-5}$. The relative errors of these solutions are measured by
comparisons with a more accurate (reference) solution (the reference
solution is obtained using double precision, $p_{fmm}=16,$ and $\epsilon
_{gmres}=10^{-6}$ ; in all cases the spherical filter of bandwidth $p_{f}=9$
was used):%
\begin{eqnarray}\label{eqErr}
&& \epsilon \left( t\right) =\frac{\left\| \mathbf{x}\left( t\right) \mathbf{-x}%
_{ref}\left( t\right) \right\| _{2}}{\max_{0<\tau <t}\left\| \mathbf{x}%
_{ref}\left( \tau \right) \mathbf{-x}_{ref}\left( 0\right) \right\| _{2}}.
\end{eqnarray}%
Here $\mathbf{x}\left( t\right) $ and $\mathbf{x}_{ref}\left( t\right) $ are
the coordinates of the surface points for the testing and reference
solutions, and $\left\| {}\right\| _{2}$ is the $L_{2}$-norm. The reason for
normalization (\ref{eqErr}) opposed to $\left\| \mathbf{x}_{ref}\right\|
_{2} $ is that $\left\| \mathbf{x}_{ref}\right\| _{2}$ for large clusters is
large compared to perturbations of the bubble surface. As a result we have a
low relative $L_{2}$-norm error, $\epsilon _{2}=\left\| \mathbf{x-x}%
_{ref}\right\| _{2}/\left\| \mathbf{x}_{ref}\right\| _{2}$, even for $%
\epsilon \sim 1$ (in our cases $\epsilon _{2}\left( t\right) /\epsilon
\left( t\right) \lesssim 10^{-2}$). In metrics (\ref{eqErr}), we obtained
for the first option $\epsilon =5.3\cdot 10^{-3}$, while for the second
option $\epsilon =2.5\cdot 10^{-4}$ (both at the 200th step).

\begin{table}[tbp]
\caption{Profiling of the FMM/GPU code (wall-clock time in seconds)}
\label{Table3}
\begin{center}
\begin{tabular}{|p{1.6cm}|p{1cm}|p{1cm}|p{1cm}|p{1cm}|}
\hline
\hline
\multirow{2}{*}{} & \multicolumn{2}{|l|}{$N=1,109,376$} & \multicolumn{2}{|l|}{$N=2,629,632$} \\ \cline{2-5}
& single & double & single  & double \\ \hline\hline
Filter & 0.12 & 0.18 & 0.25 & 0.37 \\ \hline
Surface & 0.06 & 0.08 & 0.14 & 0.20 \\ \hline
FMM DS & 0.8 & 0.8 & 2.8 & 2.8 \\ \hline
\# MVPs & 12 & 15 & 12 & 16 \\ \hline
1 MVP & 0.64 & 1.3 & 1.9 & 4.0 \\ \hline
BEM MVP & 9.4 & 21.6 & 26.8 & 70.0 \\ \hline
Time Step & 10.0 & 22.3 & 28.1 & 71.6 \\ \hline \hline
\end{tabular}%
\end{center}
\end{table}

\begin{figure}[t!]
\begin{center}
\includegraphics [scale=1]{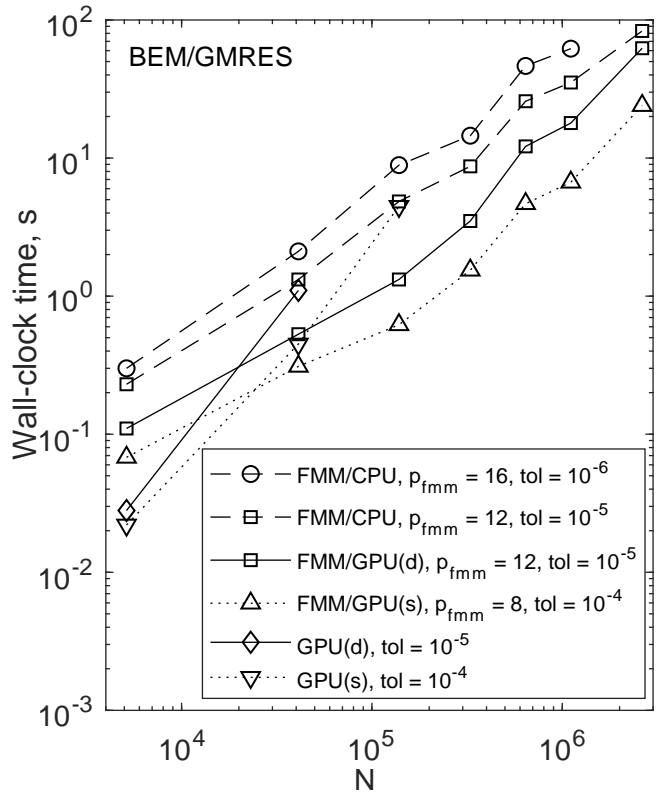}
\end{center}
\caption{The wall-clock time required for the BEM iterative solver as a
function of the number of the mesh vertices $N$ of $n_{b}\times n_{b}\times
n_{b}$ bubble clusters. The tests are performed using single and double
precision computing on GPU (marked as ``s'' and ``d''), different tolerances
for the GMRES relative residual norm $\protect\epsilon _{gmres}$ (denoted as
``tol''), and the FMM truncation numbers $p_{fmm}$ (if the FMM is employed).
The wall-clock times are measured on the same PC as in Fig. 2. }
\label{Fig11}
\end{figure}

The profile is measured for one typical time step (the 200th step from the
start; the ``warm-up'' initial steps are more expensive). The table shows
that the time for filtering applied twice for one right-hand side call is
really small compared to the time for the BEM solution (the filtering is
performed on GPU). The same applies to computations of the necessary surface
functions (briefly called ``Surface''). These functions include computations
of normals, areas, curvature, volume, and the tangential velocity and
implemented on GPU. The BEM solution requires the FMM data structure (``FMM
DS''), which computational cost normally does not exceed 10\% of the cost of
the solution. The most time is spent performing MVP. The Matlab standard
GMRES solver requires $n_{iter}+2$ MVP's for $n_{iter}$ iterations. Four
MVP's are needed to compute the singular BEM integrals, and one MVP is
needed to compute the right-hand side vector in Eq. (\ref{eq3_2}). So, the
total number of MVP's per time step of the AB6 solver is $n_{iter}+7$, which
for 2-10 GMRES iterations results in 9-17 MVP calls.

Figure \ref{Fig11} illustrates the wall-clock time required for the BEM
iterative solver. The MVP is performed using hardware and algorithmic
accelerators. Here several options with different $p_{fmm}$ and tolerance $%
\epsilon _{gmres}$ are compared. It is seen that the brute-force GPU
acceleration can be used efficiently for $N\lesssim 2\cdot 10^{4}$. For
larger $N$ combination FMM/GPU provides better performance.

\section{Conclusions}

We developed and tested an efficient numerical method, which enables
simulations of bubble systems with dynamic deformable interfaces discretized
by millions of boundary elements on personal supercomputers. It is shown
that for small and midsize ($N\lesssim 20,000$) problems GPU acceleration
alone is more efficient than the FMM/GPU acceleration, while for the
solution of large-scale problems the use of a scalable algorithm, such as
the FMM is critical. The use of GPU in the FMM brings considerable
accelerations (several times) compared to the FMM on CPU alone for the
hardware used in the present study. However, utilization of many-core CPUs
substantially reduces the effect of GPUs in the FMM, and GPU accelerations
are much smaller than that reported in \cite{Gumerov2008}. The scaling of
the algorithm obtained in this study enables estimations of the
computational time and resources for distributed computing clusters.

The algorithm is implemented and validated against simplified solutions and
solutions published in the literature. While some differences in solutions
are observed, they are explainable and do not exceed several percents
typical for solutions obtained using boundary elements. Also, validation of
the developed code was performed using different surface discretizations and
different parameter settings controlling the accuracy and stability of the
algorithm. One of the new elements implemented and tested is a shape filter,
which showed its effectiveness for mesh stabilization and efficiency
regarding performance. Profiling of the algorithm indicates that the most
time is spent on MVPs, while the overheads related to filtering, generation
of the data structure, etc. are reasonably small. It is interesting that
substantial reduction of the accuracy of computations (single precision GPU)
brings 2x accelerations, while the overall accuracy and stability are still
acceptable (the errors are much smaller than the discretization and other
BEM errors). The code can be used in many studies related to bubble
dynamics, and such applications are envisioned in future.

\end{document}